\documentclass[aps,prb,reprint, superscriptaddress,nofootinbib
]{revtex4-2}

\usepackage{graphicx}
\usepackage{dcolumn}
\usepackage{bm}
\usepackage{hyperref}
\usepackage{color}
\usepackage{eufrak}
\usepackage{amsmath}
\usepackage{amsfonts}
\usepackage{braket}

\begin{document}

\preprint{APS/123-QED}

\title{Hamiltonian Reconstruction: the Correlation Matrix and Incomplete Operator Bases}
\author{Lucas~Z.~Brito}
\affiliation{Department of Physics, Brown University, Providence, Rhode Island 02912-1843, USA}
\author{Stephen~Carr}
    \affiliation{Brown Theoretical Physics Center, Brown University, Providence, Rhode Island 02912-1843, USA}
\author{J.~Alexander~Jacoby}
    \affiliation{Department of Physics, Princeton University, Princeton New Jersey 08544, USA}
    \affiliation{Department of Physics, Brown University, Providence, Rhode Island 02912-1843, USA}
\author{J.~B.~Marston}%
    \affiliation{Brown Theoretical Physics Center, Brown University, Providence, Rhode Island 02912-1843, USA}
    \affiliation{Department of Physics, Brown University, Providence, Rhode Island 02912-1843, USA}

\date{\today}

\begin{abstract}
We explore the robustness of the correlation matrix Hamiltonian reconstruction technique with respect to the choice of operator basis, studying the effects of bases that are \textit{undercomplete} and \textit{overcomplete}---too few or too many operators respectively. An approximation scheme for reconstructing from an undercomplete basis is proposed and performed numerically on select models. We discuss the confounding effects of conserved quantities and symmetries on reconstruction attempts. We apply these considerations to a variety of one-dimensional systems in zero- and finite-temperature regimes.
\end{abstract}

\maketitle

\section{Introduction}\label{sec:introduction}
Reconstructing a Hamiltonian from a single measurable eigenstate has been a subject of recent interest in the condensed matter physics community \cite{garrison,qi,zoller,jacoby, turkeshi-parent-hamiltonians, bairey, kim, zhu, Peschel_2003, Cao_2020, Hou_2020, nandy2023reconstructing, turkeshi-jastrow-gutzwiller, kim, turkeshi-parent-hamiltonians, giudici, zhang2020}.
The reconstruction is done by way of a correlation matrix, whose elements are the expectation values of all pairs of physical observables.
The correlation matrix approach is more direct than the traditional study of quantum systems, which usually rely on matching the temperature or field dependence of electronic observables to simplified theoretical models~\cite{jacoby}. Unsurprisingly, the advantages of the single-eigenstate reconstruction comes with a severe drawback: it is hard task for a general system. The central obstacle is that, while this method is guaranteed to work when the full set of covariances between observables are obtained, in most cases only a portion of the full space of relevant physical observables are measurable.

Practical examples of this subspace problem are numerous. Often one cannot measure external couplings, account for relativistic effects, accurately sample over all material impurities, or resolve a small but non-zero two-point response from noise. More abstractly, if the Hamiltonian terms which couple to less important (incidental) set of degrees of freedom (DOFs) are small~\cite{large_norm_note}.
compared to those terms which couple to the more important DOFs, it is a decent approximation to simply trace out the incidental DOFs.
Similarly, if some part of the Hamiltonian which acts on the important DOFs is small and hard to measure or treat in a given model, one can often throw it away without harming the final result. Such approximations are a common theme across all branches of physics. In this context, the goal of the present work is to understand how neglecting such incidental degrees of freedom might hamper attempts to reconstruct a Hamiltonian via the correlation matrix method in an experimentally realistic setting.

\begin{figure}[ht!]
    \centering
    \includegraphics[width=\linewidth]{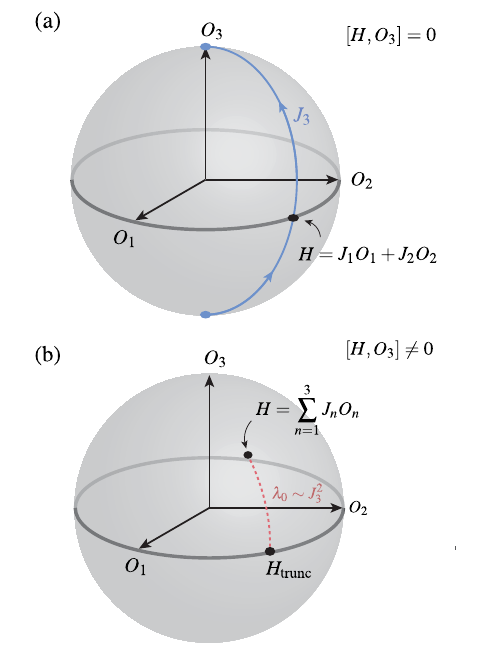}
    \caption{Schematic of the subspace problem in correlation matrix reconstruction. $\{O_n\}$ span the space of operators on the relevant sector of Hilbert space. The Hamiltonian is only defined up to a 
    positive-definite overall constant and we represent the sector as a sphere.
    (a) When the operator space is larger than necessary, the presence of a 
    conserved quantity $O_3$ causes the reconstructed Hamiltonian to be a linear combination of $O_3$ and the true $H$.
    (b) When working with an undercomplete basis, the correlation matrix approach cannot provide a perfect match to the measured state (the minimal eigenvalue $\lambda_0$ is proportional to the square of the largest truncated operator $J_3$).
    }
    \label{fig:intro}
\end{figure}

What does an experimentally realistic setting mean in the context of Hamiltonian reconstruction? The correlation matrix procedure has not yet (to our knowledge) been applied to a real system. Our interpretation is as follows: the scientist should know the basic inputs of the model---definitely a spatial symmetry group and probably also an internal symmetry group---from which they have derived a good understanding of the local DOFs. Essentially, we assume a situation where we already know of some convenient spatial tensor factorization for the Hilbert space and the global conserved quantities. Thus, we can select a basis of likely candidates for physically appropriate operators from which to reconstruct the Hamiltonian. The point of our ``incidental'' degrees of freedom is, here, to explain that we are likely to select a basis which is \textit{almost} complete, in the sense that the (appropriately normalized) operators we are neglecting appear in the Hamiltonian only with small coefficients. That is, we are likely to have addressed all of the most important physics in the chosen operator basis, but not necessarily everything.

To test if the correlation matrix is still applicable in this more realistic context, we will start with a complete basis of operators, reconstruct a Hamiltonian using the correlation matrix technique of Ref. \cite{qi}, and then study the limitations of the technique as it is pushed in opposing directions: that of an increasingly smaller operator basis, and that of an operator basis enlarged by additional operators.
A general overview of the problems caused by an enlarged or truncated basis in provided in Fig.~\ref{fig:intro}(a) and (b), respectively, for a Hilbert space with three operators.
In Section \ref{sec:methods} we review some basic features of the correlation matrix technique and the models we will utilize to study it. In Section \ref{sec:results} we determine the effects of decimating operators from a complete basis, arguing that despite incomplete knowledge of the operators in the Hamiltonian, some important features of the missing operators may still be gleaned from the correlation matrix. This effect is then demonstrated in two ways: firstly using a procedure analogous to the freezing out of the kinetic degrees of freedom in the Mott limit of a Hubbard model, in which higher order spin exchanges contribute with diminishing significance to the physics of a spin chain (Section \ref{sec:su2-dmrg}), and secondly by making use of a fairly general, \textit{long-range}-coupling translationally invariant spin model on a ring (Section \ref{sec:spin-rings}). Also in Section \ref{sec:results} we discuss the consequences of augmenting a complete basis on the spectrum of the correlation matrix jointly with the effects of the system's conserved quantities (Section \ref{sec:overcomplete}), and explore thermal effects on the reconstruction procedure (Section \ref{sec:thermal-results}). Lastly, in Section ${\rm VI}$, we conclude our study, address potential limitations, and present a number of open questions and opportunities for further work.

\section{Methods}\label{sec:methods}
\subsection{The Quantum Correlation Matrix}\label{sec:cm}
A Hamiltonian reconstruction can be performed reliably if the following criteria hold:
\begin{enumerate}
    \item The system Hamiltonian can be written as
    \begin{equation}
        H = \sum_{i}\gamma_{i}O_{i} = \vec{\gamma}_H\cdot\vec{O},
    \end{equation}
    where $\left\{O_{i}\right\}$ is a \textit{physically complete} operator basis, in the sense that it contains all operators acting nontrivially on this sector of Hilbert space.
    \item We are equipped with a state $\rho = \left(1-\epsilon\right)\left|E_{n}\right>\left<E_{n}\right|+\epsilon \varsigma$ with $\epsilon \ll 1$. Note that we assume $\varsigma \in \mathcal{D}\left(\mathcal{H}\right)$ with $\mathcal{D}\left(\mathcal{H}\right)$ the set of density operators on the system Hilbert space, $\mathcal{H}$. That is, $\rho$ is almost an eigenstate of $H$ and is associated to the state $\left|E_{n}\right>\in \mathcal{H}$ up to some contamination by $\varsigma$. A particularly important case explored in Section \ref{sec:thermal-results} is that in which $n=0$ ($\left|E_{n=0}\right>$ is the ground state) and $\varsigma$ represents thermal effects, corresponding to the experimentally realistic scenario of a large but finite inverse temperature $\beta = \left(k_\mathrm{B}T\right)^{-1}$.
    \item $H$ must be roughly local and either possesses very few conserved quantities of any degree of locality or admits states $\rho$ that are translationally invariant. Of particular interest is the case in which $H$ is translationally invariant but only approximately local.
    Our notion of locality for both the conserved quantities and the Hamiltonian is that of, e.g., Ref. \cite{hastings}; in particular, by approximately local we mean that the operator norm of each term in the Hamiltonian is at worst asymptotically vanishing as a function of site separation.
\end{enumerate}

The tool which allows us to perform such a reconstruction is the (quantum) correlation matrix. The quantum correlation matrix was introduced as a tool for Hamiltonian reconstruction in Ref. \cite{qi}, and it is defined as
\begin{equation}
    \mathcal{M}_{ij}^{\rho} = \frac{1}{2}\left<\left\{O_{i},O_{j}\right\}\right>_{\rho}-\left<O_{i}\right>_{\rho}\left<O _{j}\right>_{\rho},
    \label{eq:cormat_herm}
\end{equation}
where the expectation value is taken as $\left<O\right>_{\rho} = {\rm Tr}\left[O\rho\right]$. 

We label the eigenvalues of the matrix $\mathcal{M}^{\rho}$ as $\lambda_i$, and henceforth refer to this set of eigenvalues as the \textit{correlation spectrum}.
The correlation spectrum also defines a set of operators $\Gamma = \{\Gamma_i\}$ which carry no quantum correlations between themselves (that is, their connected correlators vanish). These operators are easily generated from the eigenvectors $\vec{\gamma}_i$ of $\mathcal{M}^\rho$ by $\Gamma_{i} = \sum_{k}\gamma_{i,\, k} O_{k} = \vec{\gamma}_{i}\cdot\vec{O}$.
If the operators $\Gamma$ are known, the correlation spectrum can also be obtained by considering $\mathcal{M}^\rho$ in this diagonalized basis, e.g.,
\begin{equation}
\lambda_{i} = \left<\Gamma_{i}^{2}\right>_\rho -\left<\Gamma_{i}\right>_\rho^{2} \equiv \sigma^2\left(\Gamma_{i} \right),
\end{equation}
where the operation $\sigma^2$ will be used for shorthand for the calculation of a diagonal element of $\mathcal{M}^\rho$. We occasionally write $\sigma^2(\Gamma_i)$ and omit $\rho$ when the state in question is unambiguous. We also write $\text{cov}(A,B)=\left< A B\right> - \left<A\right>\left<B\right>$ to denote the covariance  of two operators $A$, $B$.
It is often useful to normalize the $\vec{\gamma}_{i}$ so that we can think about $\left\{\vec{\gamma}_{i}\right\}$ as spanning the vector space of physical operators, which also induces a useful norm for operator similarity (namely, the standard inner product of two such coefficient vectors in the $\left\{O_{i}\right\}$ basis) provided the operators themselves are also suitably normalized. In what follows we will work with $\rho$ a pure state ($\epsilon=0$) before returning to the case of a thermal density matrix in Section \ref{sec:thermal-results}. 

If the correlation matrix is equipped with a complete operator basis, there will always be an element of the correlation spectrum $\lambda_{0} = 0 $ associated with an operator $\Gamma_{0} $ that is proportional to $H$. The Hamiltonian will only be reconstructed up to a positive definite multiplicative factor and up to linear contributions from additional operators which commute with the density matrix. Notably, if these commuting operators are in the span of the complete operator basis and linearly independent of the Hamiltonian, additional zeros will manifest in the correlation spectrum. We explore this case in Section \ref{sec:overcomplete}.

\subsection{Models}\label{sec:models}

Our study of the correlation matrix will focus on two different classes of 1D spin models: those with only nearest-neighbor coupling but large spin powers, and those with long-range coupling but only first-order spin powers.
The latter case is a well studied topic, while the former is not as commonly considered. Nonetheless, they provide complementary model Hamiltonians with many degrees of freedom for rigorous numerical analysis of the correlation matrix approach.

We also consider a simple, non-interacting electronic tight-binding model which can be solved in the single-particle limit. This model admits fast exact diagonalization, and allows us to study the effect of disorder and long-range couplings with a much larger dataset than those of the spin chains, which are limited in size by the computational complexity of their associated techniques. The code which generates the models presented here, as well as the results in the next section, is freely available online~\cite{github}.

\subsubsection{$SU(2)$ chains with spin $S > 1/2$}\label{sec:su2-dmrg}
The isotropic $SU(2)$ spin-chain constrained to nearest-neighbor couplings consists of exactly $2S-1$ independent terms for spin $S$, and takes the form
\begin{equation}
\label{eq:H_su2_iso}
    H^{I}_S = \sum_{p=1}^{2S} \sum_{\braket{i,j}} J^{(p)} (\vec{S}_i \cdot \vec{S}_j)^p
\end{equation}
for coupling terms $J^{(p)}$. Unlike the models that follow, we allow for higher-order interactions \smash{$(\vec{S}_i\cdot \vec{S}_j)^p$} for $p=1, \dots, 2S$ and restrict to nearest-neighbor interaction terms $\langle i,j\rangle$. We also assume the coupling terms are independent of the position in the chain $i$, but stipulate open boundary conditions such that this model is not translationally invariant. We solve these non-periodic chains using the density matrix renormalization group (DMRG) method on a matrix product operator~\cite{baker}.
To construct the correlation matrix based on these assumptions, the following complete operator basis is suitable: 
\begin{align*}
   \left\{ O_p \equiv  \sum_{i=1}^{N-1}(\vec{S}_i\cdot\vec{S}_{i+1})^p , \quad
        p=1,\dots,S-1 \right\},
\end{align*}
where $N$ is the length of the chain. We will analyze the quality of the reconstructed $H$ for this class of models by truncating operators of specific powers $p$. This is in contrast with the spatial truncation implemented in all other models, which is based on the distance between sites $i$ and $j$.

\begin{figure}
    \centering
    \includegraphics[width=\linewidth]{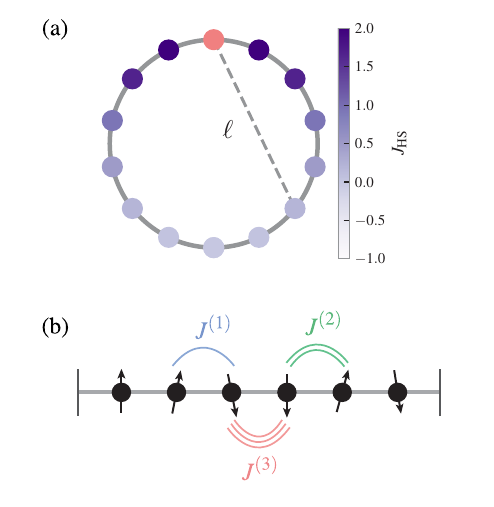}
    \caption{
    Schematic of the models used in this work. (a) A $L=14$ spin chain with periodic boundary conditions with color intensity indicating the Haldane-Shastry interaction $J_{\mathrm{HS}}(\Delta r)$ with the red site.
    (b) A finite SU(2) spin chain of Eq. \ref{eq:H_su2_iso}. The number of lines indicates the power of the interaction $\vec{S}_i\cdot \vec{S}_j$.
    }
    \label{fig:model-schematics}
\end{figure} 

\subsubsection{Translationally invariant spin-rings}\label{sec:spin-rings}
We further consider SU(2) spin chains with full translational invariance. Such systems take the form 
\begin{equation}
    H = \sum_{i>j} f\left(|i - j|\right)\vec{S}_i\cdot \vec{S}_j
    \label{eq:translational_inv_Ham}
\end{equation}
and admit as a complete basis the following translationally-averaged basis:
\begin{equation}
    \left\{ O_{i} \equiv \sum_{j}\vec{S}_i \cdot \vec{S}_{i+j} ,
    \quad i=1, \dots , N \backslash 2 \right\},
    \label{eq:operator-basis-spin-ring}
\end{equation}
where $\backslash$ denotes integer division. The choice of how to sum the indices will change the operator basis materially, and a good choice for a complete basis will depend upon the direct lattice geometry. To induce translational invariance while maintaining a finite system, we consider an $N$-site lattice arranged in a ring so that site $i+N$ is identified with site $i$, henceforth referred to as \textit{spin rings}. As is explained in Appendix \ref{app:rings}, such a geometry induces a reduction in the size of a suitable operator basis such that operators only need to be included up to separation $N\backslash 2$, meaning that the largest considered separation is the diametrically opposing point on the ring. To more compactly handle the ring geometry, it is useful to introduce the following function:
\begin{gather}
    \delta_{N}(i, j) =
    \begin{cases}
        \left|a-b\right|, & \textrm{if } \left|a-b\right|\leq \left|b+N-a\right|\\
        \left|b+N-a\right|, & \textrm{if } \left|b+N-a\right|< \left|a-b\right| 
    \end{cases}\\
    a = \operatorname{max} \left(i,j\right), \quad b = \operatorname{min}  \left(i,j\right),
    \nonumber
    \label{eq:distfunction}
\end{gather}
which measures the distance (in units of the lattice spacing) of site $i$ from site $j$ on the ring.

In this operator basis, the Hamiltonian takes the form
\begin{align*}
    H &= \sum_{j>i}^{N}f\left(\delta_{N}(i,j)\right)\, \vec{S}_{i}\cdot\vec{S}_{j} \nonumber \\
    &=\sum_{i}f\left(i\right)O_{i} . 
    \label{eq:trans_inv_ham}
\end{align*}
where in what follows we will consider power-law decaying couplings of the form $f(x) = x^{-\delta}$. The groundstates of these systems are calculated with the use of exact diagonalization schemes with the exception of the Haldane-Shastry Hamiltonian, which is described in the following section and possesses a known exact solution.

Spin rings of the above form possess a global $SU(2)$ symmetry and an associated conserved quantity in the form of the total spin:
\begin{equation}\label{eq:s_tot}
    S_{\mathrm{tot}} = \left(\sum_i^N \vec{S}_i\right)^2
        = \sum_{i,j}^N \vec{S}_i\cdot\vec{S}_j,
\end{equation}
which, in the $\{\vec{S}_i\cdot\vec{S}_j\}$ basis, manifests as a vector $\vec{1} = (1, 1, \dots, 1)$. Note that, compared to the form of the exchange couplings $f(x)$ we are considering, this is a highly ``global" quantity: the support of $S_{\mathrm{tot}}$ is the entire spin chain, a fact which, as we will demonstrate with the Haldane-Shastry model, becomes useful in discerning between the correlation matrix eigenvectors corresponding to this conserved quantity and $H$.

\subsubsection{The Haldane-Shastry Model}\label{sec:hs}
The Haldane-Shastry model \cite{haldane, shastry} is a special case of the preceding class of spin ring models consisting of an $N$-site periodic spin chain possessing the following Hamiltonian: 
\begin{gather*}
    H_{\mathrm{HS}} = \sum_{n<m} J_{\textrm{HS}}(m - n)\;
        \vec{S}_n\cdot\vec{S}_m,\\
        J_{\textrm{HS}}(x) = \frac{J\pi^2}{N^2\sin^2(x\pi/N)}
\end{gather*}
for $J>0$. The factor $1/\sin^2((m- n)\pi/N)$ in the coupling strength is equal to the square of the chord length $\ell$ extending from site $m$ to site $n$ (Fig. \ref{fig:model-schematics}a), consistent with the interpretation of sites on a physical ring coupled by an inverse-square law. The Haldane-Shastry Hamiltonian is a suitable system for our study because it is a physically realistic long-ranged Hamiltonian possessing an exact solution, the Gutzwiller projected wavefunction \cite{gutzwiller,gros}, permitting a direct calculation of the correlation matrix. The Gutzwiller projected wavefunction also possesses a closed-form two-point correlator \cite{gebhard} in the thermodynamic limit; for our purposes, however, it suffices to calculate the correlators directly.

Both the Haldane-Shastry Hamiltonian and the Gutzwiller projected wavefunction are translationally invariant, permitting the same reduction in the operator basis described in the preceding section. Likewise, the ring geometry of the system further reduces the number of required operators, as is discussed in Appendix \ref{app:rings}, such that the relevant operator coefficients are $J_{\mathrm{HS}}(j)$ for $j= 1,\dots, N\backslash 2$. Also as with the preceding models, the Haldane-Shastry model possesses an SU(2) symmetry and associated conserved quantity $S_\mathrm{tot}$. While the exchange coupling $J_\mathrm{HS}(x)$ is not of the form $x^{-\delta}$, it remains local, such that the relative globality of $S_\mathrm{tot}$ will continue to provide a useful feature to distinguish the reconstructed Hamiltonian and conserved quantity correlation matrix eigenvectors.

\subsubsection{Single-particle tight-binding model}
Finally, we will also make use of simple, single-particle one-dimensional tight-binding model for a ring of atoms.
Here we consider the electronic kinetic degrees of freedom and ignore spin.
Under these assumptions, the model has the Hamiltonian:
\begin{equation}
H_\mathrm{TB} = \sum_{i<j} t_{ij} (c_i^\dagger c_j + h.c.)
\label{eq:H_tbh}
\end{equation}
where $c_i$ ($c_i^\dagger$) annihilates (creates) an electron on site $i$, $t_{ij}$ is the nearest-neighbor hopping strength, and we assume periodic boundary conditions.

Since there are no interactions, for a ring with $N$ sites this Hamiltonian's ground state is obtained by diagonalizing an $N \times N$ matrix, instead of the computationally intractable $2^N \times 2^N$ sized matrix encountered in the presence of many-body terms.

\section{Results}\label{sec:results}

As explained in the introduction, in this work we will push the boundaries of the correlation matrix technique under non-ideal choices of the operator basis.
Failures in the operator basis can take two forms, and thus we organize our results into two sections.

First, what happens when the operator basis $\{O_i\}$ does not span $H$?
This will occur often in experiments, as not all types of operators can be measured by a single technique.
For some examples, imagine only spin terms up to $S^2$ are accessible to spectroscopy of a high-spin species, or that only spatial correlations within a certain range of distances (or momenta) can be obtained by diffraction measurements.
As the full Hamiltonian cannot be spanned by the measured operators, the correlation matrix will not have a zero eigenvalue.
We refer to these challenges as those of an \textit{undercomplete} operator basis and explore them in Section \ref{sec:undercomplete}, where we provide a perturbative approximation  connecting the size of the smallest eigenvalue of $\mathcal{M}$ to the relative strength of the missing operators.

Second, what if the operator basis $\{O_i\}$ contains operators that do not appear in $H$?
We will find that, in many cases, this poses no issue.
But in certain cases, it can introduce additional zeros to the spectra of $\mathcal{M}$, which are related to the symmetries of the wavefunction.
We refer to this class of problems as arising from an \textit{overcomplete} operator basis. Although we expect the case of an overcomplete basis to occur more rarely than that of an undercomplete basis due to the few-body 
nature of typical experimental probes, we nonetheless investigate them in Section \ref{sec:overcomplete}.

\subsection{Undercomplete operator bases}\label{sec:undercomplete}
In what follows we will consider the extent to which reconstruction with the correlation matrix can be carried out when in possession of incomplete information about the system with the use of perturbation theory. This and the finite temperature results of Section \ref{sec:thermal-results} are in the same spirit as the approximate reconstruction of Sections 3.1 and 3.2 of Ref.~\cite{qi}, only here we will explicitly consider incomplete knowledge at the level of the operator basis as opposed to at the level of the correlation matrix.

To consider the effect of dropping some of the relevant operators from our basis, we partition the true Hamiltonian into two parts.
Consider a Hamiltonian which is a sum of $N$ operators:
\begin{equation}
    H = \sum_{n= 1}^N J_n O_n.
\end{equation}
Here we assume that all of the $J_n \leq 1$, as a re-scaling of $H$ wil not affect the system's ground state. 
If we keep only the first $M-1$ operators in this list, then we can write
\begin{equation}
    H = H_0 + \Delta V
\end{equation}
where $H_0 = \sum_{n=1}^{M-1} J_n O_n$, $\Delta = J_M$, and $V = O_M + \sum_{n=M+1}^N (J_n/J_M) O_n$.
The smallest eigenvalue $\lambda_0$ of the truncated correlation matrix, e.g. $\mathcal{M}$ with just the first $M-1$ rows and columns, can also be obtained by the projection of $H$ to the truncated operator basis, namely $\sigma^2(H_0)$.
Next, we write the correlation matrix in terms of two eigenvectors, one with the smallest eigenvalue within the span of the remaining operators and one with the smallest eigenvalue within the span of the truncated operators.
These eigenvectors are the projected parts of the full $H$, e.g. $H_0$ and $V$, and we then use the associativity property of $\sigma^2$ to write this $2 \times 2$ correlation matrix perturbatively in terms of $\Delta$:
\begin{equation}
    \mathcal{M}_{ij} = 
    \begin{pmatrix}
        \sigma^2(H_0) & \text{cov}(H_0,V)\\
        \text{cov}(H_0, V) & \sigma^2(V)
    \end{pmatrix}
    =
    \sigma^2(V) \begin{pmatrix}
    \Delta^2 & -\Delta \\
    -\Delta & 1 
    \end{pmatrix}.
\end{equation}
But as $\Delta = J_M < 1$, by inspection the smallest eigenvalue in the observed correlation matrix is given by the top left matrix element, e.g.:
\begin{equation}
    \lambda = J_M^2 \sigma^2(V).
\end{equation}

In the following subsections, we will investigate the validity of this result across various models.

\subsubsection{Small $J^{(p)}$ for a nearest-neighbor spin chain}
\begin{figure}
    \centering
    \includegraphics[width=\linewidth]{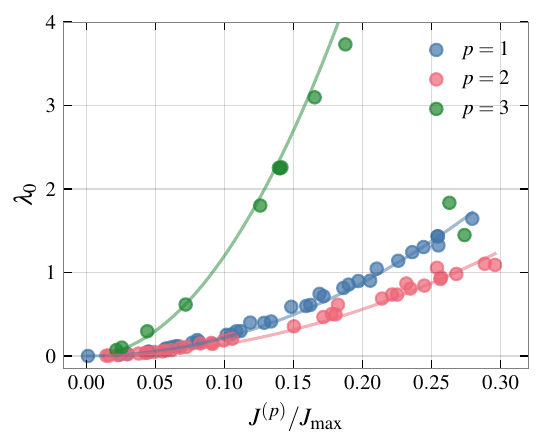}
    \caption{Dependence of smallest eigenvalue in the reduced correlation matrix on the size of the truncated $J^{(p)}$ term for randomized $SU(2)$ invariant spin-3/2 Hamiltonians (Eq.~\ref{eq:H_su2_iso} with $N = 5$). When $J^{(3)}$ is the smallest term, we only plot the cases for which $J^{(1)} < J^{(2)}$.}
    \label{fig:eig_vs_j_threehalfs}
\end{figure}

First, let us consider the isotropic SU(2) Hamiltonian of Eq. \ref{eq:H_su2_iso} with spin set to $3/2$.
For this choice of spin, the Hamiltonian is defined by the parameters $J^{(1)}$, $J^{(2)}$, and $J^{(3)}$.
The ground state is only unique up to the \textit{relative} size of these parameters, so we divide out the largest $J^{(p)}$ and consider a model with two parameters less than one and a single parameter equal to exactly 1.
To validate our perturbative approximation of the lowest eigenvalue of the correlation matrix, we randomly assign one of these three $J^{(p)}$ to be the largest ($J^{(l)} = 1$), one to be the smallest ($J^{(s)}$ randomly selected in the range $[0.0, 0.3]$) and the last to be an intermediate value ($J^{(m)}$ in the range $[0.7, 0.9]$).
After using DMRG to calculate the ground-state $\rho$ and associated correlation matrix $\mathcal{M}^\rho$, we truncate the row and column associated with the smallest parameter ($J^{(s)}$) and store the smallest eigenvalue $\lambda$ of the truncated $\mathcal{M}$.

Plotting the value of the truncated $\lambda$ against the size of the truncated $J^{(p)}$ shows a clear quadratic form (Fig.~\ref{fig:eig_vs_j_threehalfs}).
However, for each value of $p$ the quadratic form has a different curvature, showing that the value of $\sigma^2(V)$ differs for each $J^{(p)}$ truncation.
Nonetheless, it admits a universal form, and so even if we do not have any access to correlators involving $J^{(p)}$, one can estimate $J^{(p)}$ from the smallest eigenvalue.

When $J^{(3)}$ is the smallest term (green points in Fig.~\ref{fig:eig_vs_j_threehalfs}), we observe during random sampling of $J^{(1)}$ and $J^{(2)}$ that the smallest eigenvalue falls onto one of two curves with different curvature, in contrast to the universal curvature found in the $p=1$ and $p=2$ cases. These two values correspond to the ground state ordering of $\rho$, which appears to depend on if $J^{(1)} > J^{(2)}$.
Since we have constrained all parameters to be positive in this sampling, the transition is not between ferromagnetic and antiferromagnetic order, but rather due to some internal spin structure arising from competition between odd and even powers of $\mathbf{S} \cdot \mathbf{S}$ in the Hamiltonian.
We have only plotted the $J^{(1)} < J^{(2)}$ cases in Fig.~\ref{fig:eig_vs_j_threehalfs} for clarity, but note that the results for the other phase lay close to the $p=1$ curve.
Interestingly, even if $J^{(1)} < J^{(2)}$, $\rho$ can still achieve the opposite ordering if $J^{(3)}$ is sufficiently large, as evidenced by the two green dots near the much lower curvature $p=1$ curve at $J^{(3)} > 0.25$. 

\subsubsection{A long-range tight-binding Hamiltonian}
For a Hamiltonian where we assume the $J_i$ are monotonically decreasing (say, $J_i \equiv J(r_i)$, for some distance $r_i$ between sites on a lattice), we can go a step further in our perturbative treatment of $\mathcal{M}$ and write:
\begin{equation}
    \lambda(r) = J(r_\mathrm{trunc})^2 Q(r_\mathrm{trunc})
    \label{eq:lambda_smooth}
\end{equation}
where the eigenvalue and coupling are both assumed to depend smoothly on $r_\mathrm{trunc}$. The variance in the remaining operators ($\sigma^2(V)$) is now assumed to follow a smooth function of the truncated hopping distance $r_\mathrm{trunc}$, given by $Q(r_\mathrm{trunc})$.

\begin{figure}
    \centering
    \includegraphics[width=\linewidth]{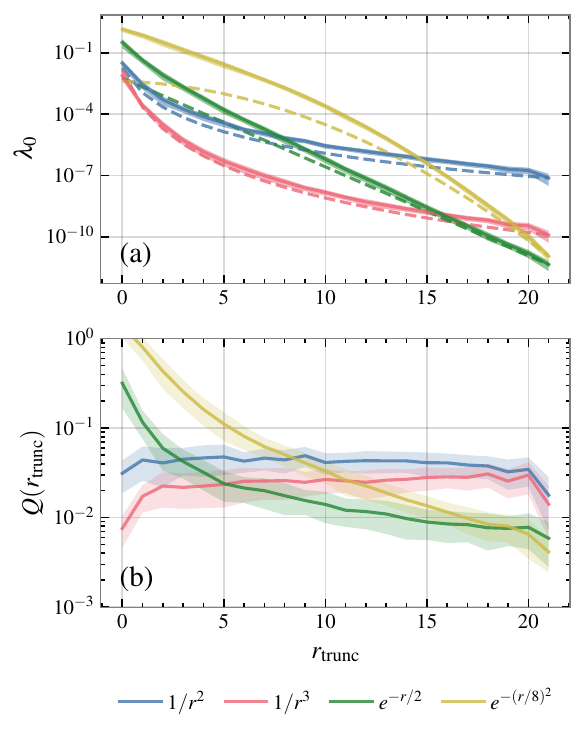}
    \caption{(a) Dependence of the smallest eigenvalue $\lambda$ of the truncated correlation matrix on the largest kept term $r$ for a noisy $N=50$ 1D tight-binding Hamiltonian for four types of coupling decay forms.
    Results are obtained by averaging over 50 random samples of the Hamiltonian at each $R$ truncation value, with the solid line indicating the mean value and the shaded region providing the range of one standard deviation.
    The dashed line gives the value of $J(r)^2$, showing that for each model, most of the variation in $\lambda(r)$ is explained by the square of the largest truncated operator parameter of the reduced basis.
    (b) Dependence of $Q(r_\mathrm{trunc}) = \lambda(r_\mathrm{trunc})/J(r_\mathrm{trunc})^2$ on $r_\mathrm{trunc}$ for simple tight-binding models of different $J(r_\mathrm{trunc})$ forms.. The solid line gives the mean value, while the shaded region gives the range of one standard deviation.}
    \label{fig:Q_fitting}
\end{figure}

We test this approximation of $\mathcal{M}$ on a simple single-particle one-dimensional tight-binding model, as defined in Eq.~\ref{eq:H_tbh}.
If we assume the hoppings depend only on the distance between sites $t_{ij} \equiv t(|i-j|)$, the model obtains complete translational symmetry and the electronic DOFs obey conservation of crystal momentum.
As we will see in the next results section, the inclusion of many additional conserved quantities (crystal momentum) will cause an equivalent number of zeros to enter the spectra of $\mathcal{M}$.
To avoid this issue, we destroy translational symmetry by adding a fixed percentage of noise to each coupling, e.g. for a given sample of the Hamiltonian we set
\begin{equation}
t_{ij} = \theta_{ij} f(|i-j|)
\label{eq:tbh_t}
\end{equation}
where $\theta_{ij}$ is a random number sampled uniformly from the range $[0,1]$ which is unique for every pair of sites.
We test the validity of Eq.~\ref{eq:lambda_smooth} for various forms of the function $f(r)$ in Fig.~\ref{fig:Q_fitting}a.
As every instance of a Hamiltonian is unique due to the randomness introduced by $\theta_{ij}$, we take $50$ samples of each functional form and report the average value and standard deviation of $\lambda$.
For most forms of $f(r)$, almost all of the variance in $\lambda$ is explained by $J(r)$ alone, as shown in Fig.~\ref{fig:Q_fitting}b.
One can, to good approximation, take $Q(r_\mathrm{trunc}) = Q_0$, a constant, for most values of $r$.
The only hopping form which showed strong dependence of $Q$ on $r_\mathrm{trunc}$ is the form with the sharpest decay, $J(r) \sim e^{-(r/8)^2}$.

Our observation of the constant prefactor $Q_0$ inspires a more robust procedure of reconstructing the parameters $J(r)$ of $H$ than simply looking at the eigenvector $\mathcal{M}$.
The procedure is as follows: artificially continue the truncation of $\mathcal{M}^\rho$, throwing out measured observables along the way. Then, estimate the Hamiltonian coupling at every measured coupling range by the obtained lowest eigenvalue, following $J(r) \propto \sqrt{\lambda_0(r)}$. 

\subsection{Overcomplete operator bases and conserved quantities}\label{sec:overcomplete}
As discussed in Section \ref{sec:cm}, the correlation matrix is diagonalized by operators $\Gamma_i$ whose eigenvalues are equal to their variances $\sigma^2(\Gamma_i)=\left<\Gamma_i^2\right>- \left<\Gamma_i\right>^2$. The nullspace of the correlation matrix thus comprises the algebra of operators that commute with the provided density matrix $\rho$. In an ideal scenario, this nullspace is of dimension one, i.e., the only operator with vanishing variance in the space spanned by the operator basis $\{O_i\}$ is the Hamiltonian $H=\vec{\gamma}_H\cdot \vec{O}$. In this case, the nullspace encodes $\vec{\gamma}_H$ up to rescaling. Indeed, one expects that the overwhelming majority of the operators that could act on a Hilbert space fall outside the sector spanned by $\{O_i\}$, such that any reasonable complete operator basis suffices to reconstruct the Hamiltonian unambiguously. However, there are cases where the nullspace is enlarged by operators other than the Hamiltonian which are hidden in the span of the selected operator basis.

Specifically, additional zeros appear in the spectrum of the correlation matrix whenever there are operators in the span of $\{O_i\}$ that possess vanishing variance and are linearly independent from the Hamiltonian. One such scenario arises when symmetries of the two-point correlators are not fully resolved by the choice of operator basis. Concretely, consider two distinct operators $O_\ell$ and $O_\ell'$ with identical two-point correlators with every other operator in the basis. If one starts with a \textit{complete} basis $\{O_1, \dots, O_\ell \}$ and enlarges it by including $O_\ell'$ to form a new basis $ \{O_1, \dots, O_\ell ,O_\ell' \}$ or $\vec{O}'$, the correlation matrix sees as a zero-variance operator both the Hamiltonian in the form $\vec{\gamma}_H = (\gamma_{H,1}, \dots, \gamma_{H,\ell}, 0)$ and the operator 
\begin{equation*}
    \vec{\gamma}'_H \cdot \vec{O}' ,\quad \vec{\gamma}' =(\gamma_{H, 1}, \cdots, 0, \gamma_{H,\ell}),
\end{equation*}
i.e., the coefficient of $O_\ell$ is taken to $O_\ell'$. These two operators are manifestly linearly independent, and the correlation matrix thus accrues an additional zero; the inclusion of $O_\ell'$ thus makes $O_\ell'$ an overcomplete basis. An example of this phenomenon is the redundancy in the translationally averaged basis explained in Appendix \ref{app:rings}, wherein operators at separations $|i-j|$ and $|i-(N-j)|$ have identical two point correlations. 

Another source of additional zeros is rooted in the notion of a correlation matrix symmetry.
This is a symmetry embedded in the correlation matrix, and is not necessarily a symmetry observed by the Hamiltonian.
We say that a Hamiltonian enjoys a symmetry if the Hamiltonian commutes with the Cartan subalgebra of the symmetry's generators. Thus we demand that any eigenstate transform according to an irreducible representation of the associated algebra \cite{georgi}.
The correlation matrix, on the other hand, views as a symmetry any operator which commutes with $\rho$, Generally, we expect there to be a very large number of operators that commute with $\rho$. Luckily, \textit{most} such operators will not lay in the span of the chosen operators, $\left\{O_{i}\right\}$, but two types of commuting operators that do lay in this span occur frequently.

First, if the system possesses a conserved quantity $J$ with $[H,J] = 0$, the operator $J$ will commute with $\rho$ and thus its variance will vanish. Notably, $J$ may generally lie in the span of a complete operator basis, such that the correlation matrix suffers from additional zeros despite being equipped with a complete basis. 
Indeed, similar findings are reported in Ref.~\cite{lian} in the context of entanglement 
Hamiltonians, and it is unsurprising that an analogous complication arises in the correlation matrix.
We observe precisely this situation observed in Sections \ref{sec:spin-rings-results} and \ref{sec:hs-results} with the total spin charge actings as the conserved quantity arising from the systems' SU(2) symmetries. In this case the symmetry is observed both by the correlation matrix and the Hamiltonian.  

Second, sometimes the ground-state wavefunction's symmetry differs from that of the Hamiltonian's symmetry. In particular, the wavefunction might possess a higher degree of symmetry than the Hamiltonian: there may be operators which are \textit{not} conserved quantities, but commute with the Hamiltonian when restricted a subspace, a form of Hilbert space fragmentation~\cite{fragmentation_note,moudgalya-review}. In such cases, sectors of Hilbert space are labelled by irreducible representations of so-called \textit{commutant algebras} \cite{moudgalya-commutant-algebras} as opposed to conventional conserved quantities of the system. Such behavior is exhibited by, for instance, the $t$--$J_z$ model in the form of a product of number operators which does not commute with $H_{t-J_z}$ but nonetheless possesses zero variance with respect to certain eigenstates of the Hamiltonian \cite{zhang1997, moudgalya-commutant-algebras}.

\subsection{Results from Translationally Invariant Spin Rings}\label{sec:spin-rings-results}
\begin{figure}
    \centering
    \includegraphics[width=\linewidth]{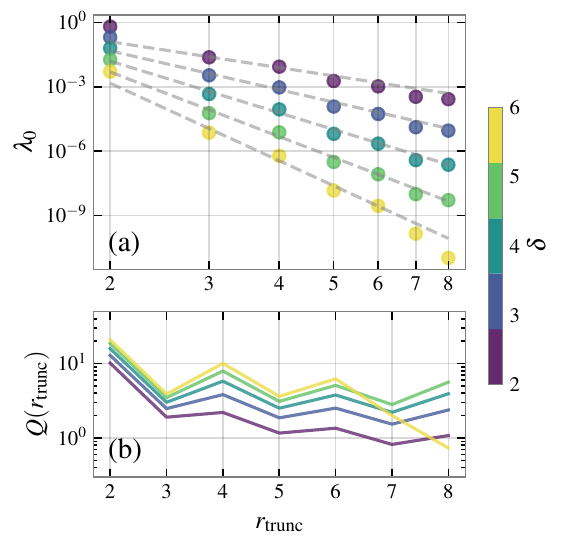}
    \caption{(a) Dependence of smallest eigenvalue $\lambda_0$ of translationally-invariant spin rings with various coupling decays $\delta$ ($N=18$). The dashed lines are functions proportional to $r_\mathrm{trunc}^{-2\delta}$. (b) Dependence of the variance $Q(R)$ of the truncated operators on the separation of the largest truncated operator. Note the relatively constant value at separations 3--7.}
    \label{fig:spin-rings-results}
\end{figure}

As in the preceding sections, we study the effects of an undercomplete basis on reconstruction by repeatedly computing and diagonalizing the correlation matrix, starting from the basis defined in Eq.~\eqref{eq:operator-basis-spin-ring} and sequentially truncating the operator acting at the largest separation. We find agreement with the result of Section \ref{sec:undercomplete}: for each coupling decay power, $\delta= 2,\dots, 6$, the smallest eigenvalue in the correlation spectrum is well-approximated by $Q_0 r_\mathrm{trunc}^{-2\delta}$ for intermediate values of $r_\mathrm{trunc}$, where $r_\mathrm{trunc}$ is the separation the largest truncated operator (Fig \ref{fig:spin-rings-results}a). The $Q_0$ form is a good approximator of the smallest eigenvalue as long as the variance of the truncated spin-spin interactions $Q(r_\mathrm{trunc})$ is roughly constant. It remains nearly constant when $r_\mathrm{trunc} > 3$, but for stronger decays ($\delta = 6$) $Q(r_\mathrm{trunc})$ begins to undergo more dramatic variations, signalling a breakdown of the $r_\mathrm{trunc}^{-2\delta}$ form (Fig. \ref{fig:spin-rings-results}b).

\subsubsection{Results from the Haldane-Shastry Hamiltonian}\label{sec:hs-results}

\begin{figure}
    \centering
    \includegraphics[width=\linewidth]{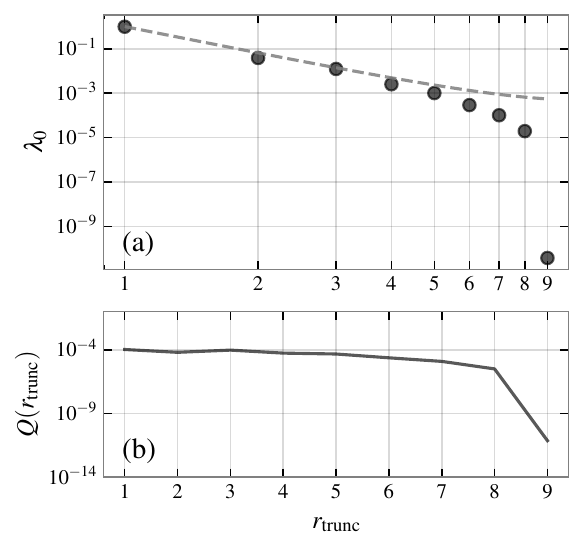}
    \caption{(a)--(b) Dependence of smallest eigenvalue $\lambda_0$ of the Haldane-Shastry correlation matrix on the number of kept operators over the magnitude of the truncated term's coefficient for an $N=21$ chain, in analogy with Fig. \ref{fig:Q_fitting}a. (c). Dependence of $Q(r_\mathrm{trunc}) = \lambda_0(r_\mathrm{trunc})/J^2_{\mathrm{HS}}(r_\mathrm{trunc})$ on number of kept operators. Note $Q(r_\mathrm{trunc})$ is approximately constant in the range 1--8.
    }
    \label{fig:hs-evals-coeffs}
\end{figure}

The reconstruction of the Haldane-Shastry Hamiltonian proceeds in a manner identical to that of the spin rings of the preceding section. Again using the operator basis of Eq.~\eqref{eq:operator-basis-spin-ring}, we find further agreement with the result of Section \ref{sec:undercomplete} in that the smallest eigenvalue of the correlation matrix is, to a good approximation, proportional to the square of the Haldane-Shastry coupling strength $J_{\mathrm{HS}}(j)$ at the smallest truncated separation (Fig. \ref{fig:hs-evals-coeffs}a--b). This is the case for intermediate separations, owing to the approximately constant value of the factor $Q(r_\mathrm{trunc})$ in this region; at larger separations, the variance in the truncated operators undergoes more dramatic variations, the lowest eigenvalue suddenly drops considerately as we approach a complete operator basis, and this approximation is no longer valid.

\begin{figure}
    \centering
    \includegraphics[width=\linewidth]{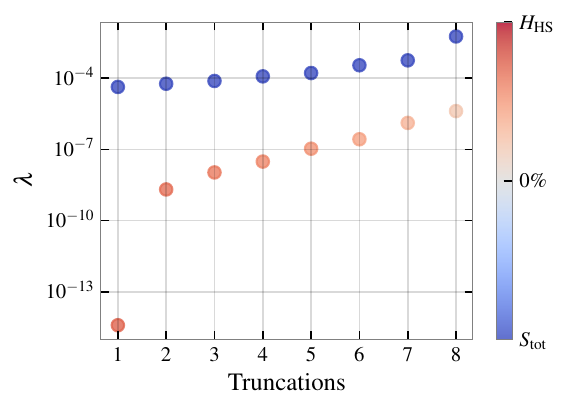}
    \caption{
    Two lowest eigenvalues of the correlation matrix of the Haldane-Shastry ground state, colored according to the bias of the corresponding eigenvector toward the Hamiltonian and the conserved quantity ($S_\mathrm{tot}$). 
    }
    \label{fig:hs-cons-quant}
\end{figure}

The presence of the conserved quantity $S_{\textrm{tot}}$ mentioned in Section \ref{sec:hs} produces an additional zero when reconstructing with a complete basis. Computing the correlation matrix with incomplete bases, however, leads to a splitting in the two lowest eigenvalues---the Haldane-Shastry coupling, being a local quantity, maintains a lower variance as operators are truncated as operators at increasing separations contribute increasingly less to it owing to the decaying nature of $J_\textrm{HS}$. The variance of the conserved quantity $S_\textrm{tot}$, on the other hand, departs further from zero as it is a global quantity; i.e., every site in the chain contributes uniformly. This can be seen in spectrum of the correlation matrix by employing the notion of operator similarity introduced in Section \ref{sec:cm}. In particular, one can measure the \textit{bias} toward either quantity by normalizing the operator similarity according to full similarity with either $H_\mathrm{HS}$ or $S_\mathrm{tot}$; this is performed in Fig. \ref{fig:hs-cons-quant} to reveal that indeed the eigenvector closest to the conserved quantity maintains a higher eigenvalue than that of the eigenvector closest to $H_\textrm{HS}$.

\subsection{Finite Temperature Results}\label{sec:thermal-results}
Understanding the accuracy of reconstruction from an impure state $\rho =(1-\epsilon) \ket{E_0}\bra{E_0} + \epsilon \varsigma$ is relevant in experiments, which always occur at finite temperature. In what follows, we will take $\epsilon \ll 1$, which corresponds to a large inverse-temperature Boltzmann weight (low, but finite temperature). Using correlators constructed from the mixed state $\rho$, the correlation matrix corresponding to an operator basis $\{O_i\}$ can be written
\begin{equation*}
    \mathcal{M}_{ij} = (1-\epsilon)\mathcal{M}_{ij}^0 + \epsilon \eta^\varsigma(O_i, O_j).
\end{equation*}
$\mathcal{M}_{ij}^0$ denotes the correlation matrix as defined in Eq.~\eqref{eq:cormat_herm} with respect to the pure state $\ket{E_0}\bra{E_0}$. The perturbative results of Section~\ref{sec:undercomplete} will continue to hold for this term, unaffected by the finite-temperature contamination. $\eta^\varsigma (O_i,O_j)$ is a matrix defined by 
\begin{align}
    \eta^\varsigma(A,B) 
        &= \langle AB \rangle_\varsigma + (1-\epsilon) \langle A \rangle_0 \langle B\rangle_0 \nonumber \\
           &\quad - (1-\epsilon) \langle A \rangle_0 \langle B \rangle_\varsigma 
           - (1-\epsilon) \langle A \rangle_\varsigma \langle B \rangle_0 \nonumber \\
           &\quad - \epsilon \langle A \rangle_\varsigma \langle B \rangle_\varsigma.
           \label{eq:eta}
\end{align}
Because the reconstructed Hamiltonian is only unique up to an overall multiplicative factor, the reconstruction procedure is unaffected by dividing $\mathcal{M}_{ij}$ by $ (1-\epsilon)$ to normalize the first term; this is equivalent to stating that only the relative contamination from $\varsigma$ is relevant to the spectrum of the correlation matrix. Based on this observation, we redefine 
\begin{equation*}
    \mathcal{M}_{ij} \rightarrow \mathcal{M}_{ij}' = \mathcal{M}_{ij}^0 + \frac{\epsilon}{1-\epsilon} \eta^\varsigma(O_i,O_j).
\end{equation*}

\begin{figure}
    \centering
    \includegraphics[width=\linewidth]{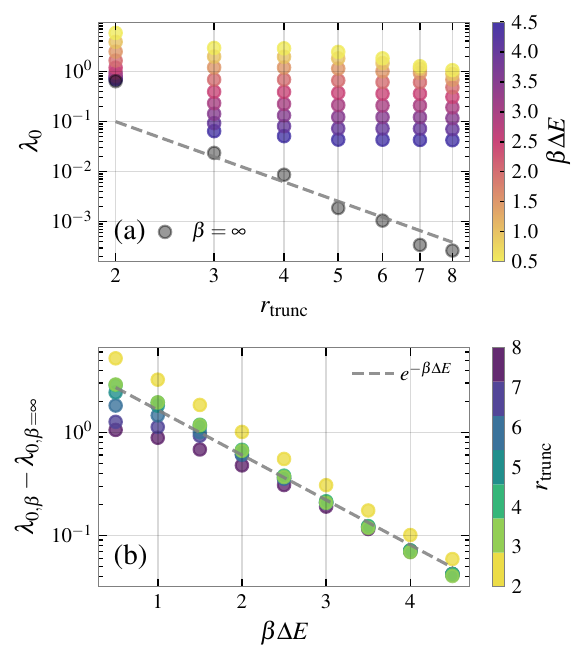}
    \caption{
    Results of finite temperature reconstruction trials for $L=18$ spin rings with periodic boundary conditions and coupling decay $\delta=2$. (a) Dependence of lowest eigenvalue of correlation matrix on number of kept operators for various coldnesses $\beta$. The dashed line represents the functional form $r_\mathrm{trunc}^{-2\delta}$ (b) Distance of lowest eigenvalues of finite temperature correlation matrices from zero-temperature lowest eigenvalue of the correlation matrix against increasing coldness.
    }
    \label{fig:thermal-results}
\end{figure}

The case of a finite temperature measurement is captured by the general form 
\begin{equation*}
    \frac{\epsilon}{1-\epsilon}
        = e^{-\beta \Delta E}, 
    \quad 
    \varsigma = \sum_{i>0} e^{-\beta (E_i - E_1)}\ket{E_i}\bra{E_i}
\end{equation*}
where $\Delta E = E_1 - E_0$ is the gap between the ground state and first excited state and where we make $\beta$ dimensionless by working in units of the gap. Thus, in analogy with Section \ref{sec:undercomplete}, we conclude that the lowest eigenvalue follows
\begin{equation*}
    \lambda(r_\mathrm{trunc}) = Q(r_\mathrm{trunc}) J^2(r_\mathrm{trunc}) + \frac{\epsilon}{1-\epsilon}\eta^\varsigma (H_0, H_0)
\end{equation*}
such that for a fixed operator basis and in the small $\epsilon$ (large $\beta$) regime, the smallest eigenvalue of the correlation matrix acquires an additive error proportional to $e^{-\beta\Delta E}$. At larger temperatures, one expects contributions from higher ($E_i$ for $i>1$) Boltzmann weights contained in $\varsigma $ to contribute to this error through $\eta^\varsigma$. At the temperature scale of the gap, however, $\eta^\varsigma(O_i,O_j)$ is approximately constant in $\epsilon$ and the asymptotic $e^{-\beta \Delta E}$ behavior is a good approximation of the finite-temperature contribution to the lowest eigenvalue (Fig. \ref{fig:thermal-results}b). 
For a fixed $\beta$, the separation from the zero-temperature lowest eigenvalue retains a dependence on the truncation separation $r_\mathrm{trunc}$ through $\eta^\varsigma(H_0, H_0)$; the separation increases with the size of the truncated operator due to the presence of the expectation values of $H_0$ and $H_0^2$, which themselves can be written as quadratic functions of $\Delta$ per Section \ref{sec:undercomplete}.

In the case of a gapless system the method accrues an error constant in $\epsilon$ and the situation evidently worsens. This is perhaps to be expected as it is common for quantum many-body methods to encounter difficulties arising from vanishing spectral gaps \cite{bursill,legeza,andersson}. In the event that the correlators of the ground state and gapless excitations are sufficiently similar, one might speculate that the correction remains small enough to retain some efficacy---this is plausible if the low-lying excitations are, for instance, described by long-wavelength Bloch-waves built from the ground state. Nevertheless the accuracy of the correlation matrix methods remains uncertain in this regime and further care in investigating this technique's use at finite temperature is needed.

\section{Conclusion}
This work has explored the correlation matrix reconstruction technique introduced by Ref. \cite{qi} as subjected to experimentally realistic, imperfect choices of operator bases. We have devised a strategy to perturbatively obtain the magnitudes of the missing terms of the Hamiltonian by leveraging the variance with respect to incomplete operator bases, and verified their accuracy using an assortment of one-dimensional models. This result suggests that even relatively local or otherwise limited probes of the many-body system's two-point correlations suffice to approximately reconstruct the system using the correlation matrix. We have additionally demonstrated that improper resolution of the wavefunction's symmetries significantly harms the reliability of the technique by enlarging the correlation matrix's nullspace, while the presence of conserved quantities similarly complicates matters even in possession of a complete basis, as does the presence of unconventional symmetries of the wavefunction (exhibited by recently-discovered classes of ergodicity-breaking systems~\cite{moudgalya-review, moudgalya-commutant-algebras}).

On the experimental front, this work suggests that the correlation matrix is a promising approach to understanding the many-body Hamiltonians of condensed matter and cold-atom systems through the usage of standard, local probes of correlations such as scattering techniques. On the theoretical front, one should be able to reconstruct the Hamiltonian corresponding to, say, a particular wavefunction ansatz, where otherwise such a search would be non-straightforward

The systems under consideration were relatively limited in scope in order to simplify our demonstration of the phenomena of interest---an extended investigation of the ideas presented in this work might make use of higher-dimensional systems, models with more diverse interactions, or a more careful treatment of the thermodynamic limit of such models. In particular, the effects of Hilbert space fragmentation on the correlation matrix technique deserve a more careful study and present a natural extension of this work.

\appendix

\begin{acknowledgements}
L. Z. B. acknowledges support from a Karen T. Romer Undergraduate Teaching and Research Award (UTRA).
J.~A.~J. would like to thank Jonah Herzog-Arbeitman for insightful conversations.
This work was supported in part by U.S. National Science Foundation Grants No. OIA-1921199 and No. OMA-1936221. J.~A.~J. was also supported U.S. National Science Foundation Grant No. DGE-2039656 for part of the duration of this work. Any opinions, findings, and conclusions or recommendations expressed in this material are those of the authors and do not necessarily reflect the views of the National Science Foundation.
\end{acknowledgements}

\section{Ring Geometries}
\label{app:rings}
In cases of fully translationally invariant Hamiltonians on a ring, the basis of independent operators is substantially diminished. It should be noted that a similar operator basis decimation is required for any systems with toroidal sub-manifolds, and not just ring geometries. However, the ring basis will provide all necessary tools to understand how this process applies in nearly arbitrary geometries.

As is alluded to in the definitions for Eq. \eqref{eq:operator-basis-spin-ring} by the phrase ``full translational invariance,'' our Hamiltonian and wavefunction, $\left|\psi\right>$, must \textit{both} be invariant under translations. This condition can only apply in the case of a ring-like geometry or the thermodynamic limit. Finite chains with hard boundaries inherently contain finite size effects which break the local translational invariance of the wavefunction. Explicitly, within a finite chain we always find that the reduced density matrix of DOFs at the edges are different than those in the deep bulk, even if only slightly. In many cases these effects are small or can be made small if one is willing to trace out some of the edge DOFs. Thus ring-like boundary conditions are often a good approximation. In this regime, using the translationally averaged basis of Eq. \eqref{eq:operator-basis-spin-ring} is desirable; however, directly applying this averaged basis will lead to a overcomplete basis.

The reason the averaged basis is overcomplete is the following. Let us consider an operator acting at separation $r = N - 1$ on a ring with $N$ sites. The ring geometry of the system forces that operator be identified with that same operator acting at separation $r=1$. More generally, if $O(|i-j|)$ is an operator that depends on $|i-j|$, a ring lattice requires $O(|i -j|) = O(|i - (N- j)|)$, as past the diametrically opposed site of the lattice---or for odd-sized lattices, at the first site past the diameter---the separation begins to ``loop around'' such that the actual site separation is smaller than $|i-j|$. For the purposes of reconstruction, this has the effect of reducing the size of a suitable basis to $N \backslash 2$, where $ \backslash $ denotes integer division. Eigenstates of $H$ are simultaneous eigenstates of a new Hamiltonian obtained by exchanging operators at separations $|i-j|$ and $|i-(N-j)|$; consequently, the nullspace of the correlation matrix increases in dimensionality for each kept operator acting at separation greater than $N\backslash 2$. The ``reflection symmetry'' with respect to $i\leftrightarrow j$ of $O(|i-j|)$ is a necessary but insufficient condition for this reduction in the operator basis to occur; such a symmetry justifies the usage of a basis labelled by separation alone (that is, by $|i-j|$ as opposed to $i-j$), but a toroidal system geometry is additionally necessary to permit this halving of the operator basis.

\nocite{*}

\bibliographystyle{apsrev4-1}
\bibliography{Ref}

\begin{thebibliography}{34}
\bibitem [{lar()}]{large_norm_note}%
  \BibitemOpen
  \href@noop {} {}\bibinfo {note} {Extremely large norms can also make certain
  DOFs appear unimportant. This is because a sufficiently strong interaction
  acts like a constraint on the low energy modes. A classic example of this
  principle is the Hubbard model in the Mott limit, where strong on-site
  interactions freeze out the kinetic degrees of freedom entirely. The only
  degrees of freedom left are the spin degrees of freedom, which become the
  ``important'' ones.}\BibitemShut {Stop}%
\bibitem [{fra()}]{fragmentation_note}%
  \BibitemOpen
  \href@noop {} {}\bibinfo {note} {This is perhaps an abuse of terminology as
  some authors might additionally stipulate the presence of exponentially many
  such sectors and a non-integrable Hamiltonian. Here we require neither
  exponential multiplicity nor non-integrability, as the wavefunction and thus
  the correlation matrix will view such quantities as bearing vanishing
  variance regardless of these two features.}\BibitemShut {Stop}%
\end{thebibliography}

\end{document}